FRONT MATTER

## Title

# Rapid and high-sensitive NV-based microwave field imaging via digital lock-in amplification for on-chip microstrip diagnostics

## Authors

Zijin Fu[1,2], Yanjie Liu[1,3], Hongliang Wu[1,3], Yuchen Han[1,2], Zhengtao Wang[1,2], Haolin Li[1,2], Dezhi Zheng[4,5], Bo Zhang[1,2,4,5,6*] and Jun Zhang[4,5,7,*]

**Affiliations**

[1]Key Laboratory of Advanced Optoelectronic Quantum Architecture and Measurements of Ministry of Education, Center for Interdisciplinary Science of Optical Quantum and NEMS Integration, School of Physics, Beijing Institute of Technology, Beijing 100081, China

[2]Center for Photonic Quantum Precision Measurement, Advanced Research Institute of Multidisciplinary Science, Beijing Institute of Technology, Beijing 100081, China

[3]School of Integrated Circuits and Electronics, Beijing Institute of Technology, Beijing 100081, China

[4]State Key Laboratory of Environment Characteristics and Effects for Near-space, Beijing Institute of Technology, Beijing 100081, China

[5]MIIT Key Laboratory of Complex-field Intelligent Sensing, Beijing Institute of Technology, Beijing 100081, China

[6]Center for Quantum Technology Research and Key Laboratory of Advanced Optoelectronic Quantum Architecture and Measurements (MOE), Beijing Institute of Technology, Beijing 100081, China

[7]State Key Laboratory of CNS/ATM, Beijing Institute of Technology, Beijing 100081, China

* Correspondence: bozhang_quantum@bit.edu.cn (B. Z.), zhjun@bit.edu.cn (J. Z.).

## Abstract

High-resolution, high-sensitivity microwave (MW) magnetic field imaging is indispensable for non-destructive integrated circuit (IC) testing, radio-frequency device characterization, and spintronic research. Yet, the practical utility of these techniques is severely constrained by the pervasive challenge of isolating weak magnetic signatures from intense optical and electronic noise, which fundamentally limits both acquisition speed and detection sensitivity. Here, we overcome this barrier by introducing a wide-field imaging scheme based on an ensemble of diamond nitrogen-vacancy (NV) centers, synergistically combined with digital lock-in amplification (DLA). By exploiting digital demodulation, the DLA precisely extracts the MW-field response at a specific modulation frequency from background noise (e.g., laser intensity fluctuations), dramatically improving the signal-to-noise ratio (SNR). Consequently, our system attains a magnetic field sensitivity of 126 $nT/Hz^{1/2}$. Critically, the unprecedented SNR permits a pixel dwell time of under one millisecond, allowing full-field images to be acquired within seconds-more than an order of magnitude faster than state-of-the-art NV-based wide-field techniques. This combination of speed, sensitivity, and micron-scale spatial resolution (1.6 μm) paves the way for quasi-real-time, non-invasive diagnostics of dynamic MW devices and integrated circuits.

## MAIN TEXT

## INTRODUCTION

The relentless progression of wireless communications[1], the Internet of Things (IoT), and high-performance computing is driving radio-frequency and microwave (RF/MW) integrated circuits (ICs) toward ever-higher operating frequencies[2,3], power densities[4], and integration levels[5]. This trajectory imposes stringent demands on non-destructive characterization techniques, particularly for the quantitative, micrometer-scale imaging of internal MW magnetic fields—a capability essential for understanding signal integrity, localizing electromagnetic interference (EMI) sources, and diagnosing thermal hotspots and latent defects[6–8]. Conventional near-field probe scanning, though widely used, is inherently limited by poor spatial resolution (on the order of millimeters) and intrusive coupling with the device under test, rendering it ill-suited for next-generation nanoscale electronics[9].

Diamond-based quantum sensing using nitrogen-vacancy (NV) centers[10] has emerged as a compelling alternative, offering atomic-scale spatial resolution, room-temperature operability, and sensitivity to magnetic fields[11–14], stress field[15,16], and temperature[17,18]. Through optically detected magnetic resonance (ODMR)[19,20], NV ensembles enable wide-field MW field mapping with sub-micrometer potential[21,22]. In practice, however, the fluorescence contrast in such measurements is intrinsically weak and vulnerable to optical shot noise, laser intensity jitter, and camera readout noise. As a result, conventional wide-field NV imaging relies on extensive signal averaging over hundreds to thousands of frames, leading to acquisition times of minutes to hours a fundamental bottleneck that precludes dynamic observation and high-throughput screening. Breaking this trade-off among sensitivity, spatial resolution, and temporal throughput remains a central challenge in the field.

Here, we introduce a paradigm shift to overcome this limitation by integrating digital lock-in amplification (DLA) with wide-field NV magnetometry[23], as shown in Fig.1(a). Rather than averaging raw ODMR signal[24], we frequency--modulate the MW source and synchronously demodulate the camera--detected fluorescence signal using a custom--developed DLA algorithm. This DLA scheme selectively extracts the MW-field response coherent with the modulation frequency while effectively rejecting uncorrelated background noise—including laser fluctuations and sensor readout noise—within a single or very few modulation cycles. Our approach achieves a magnetic field sensitivity of 126 nT/Hz$^{1/2}$, a spatial resolution of 1.6 μm(the resolution analysis is provided in Supplementary Material S1). and, crucially, a full-field imaging time of just seconds, representing a two-order-of-magnitude acceleration compared to conventional NV-based wide-field microscopy. We designed two experimental schemes: (1) measuring a microstrip line with an applied MW signal, and (2) measuring a microstrip line without an applied signal (requiring a horn antenna to provide the MW signal). We also validate this method through two complementary experiments: (1) imaging the MW field distribution above a driven microstrip line, and (2) fault-diagnosis mapping of an intentionally broken circuit. These results establish DLA-enhanced NV imaging as a practical, high-speed, and high-fidelity tool for non-destructive testing and failure analysis of advanced RF devices.

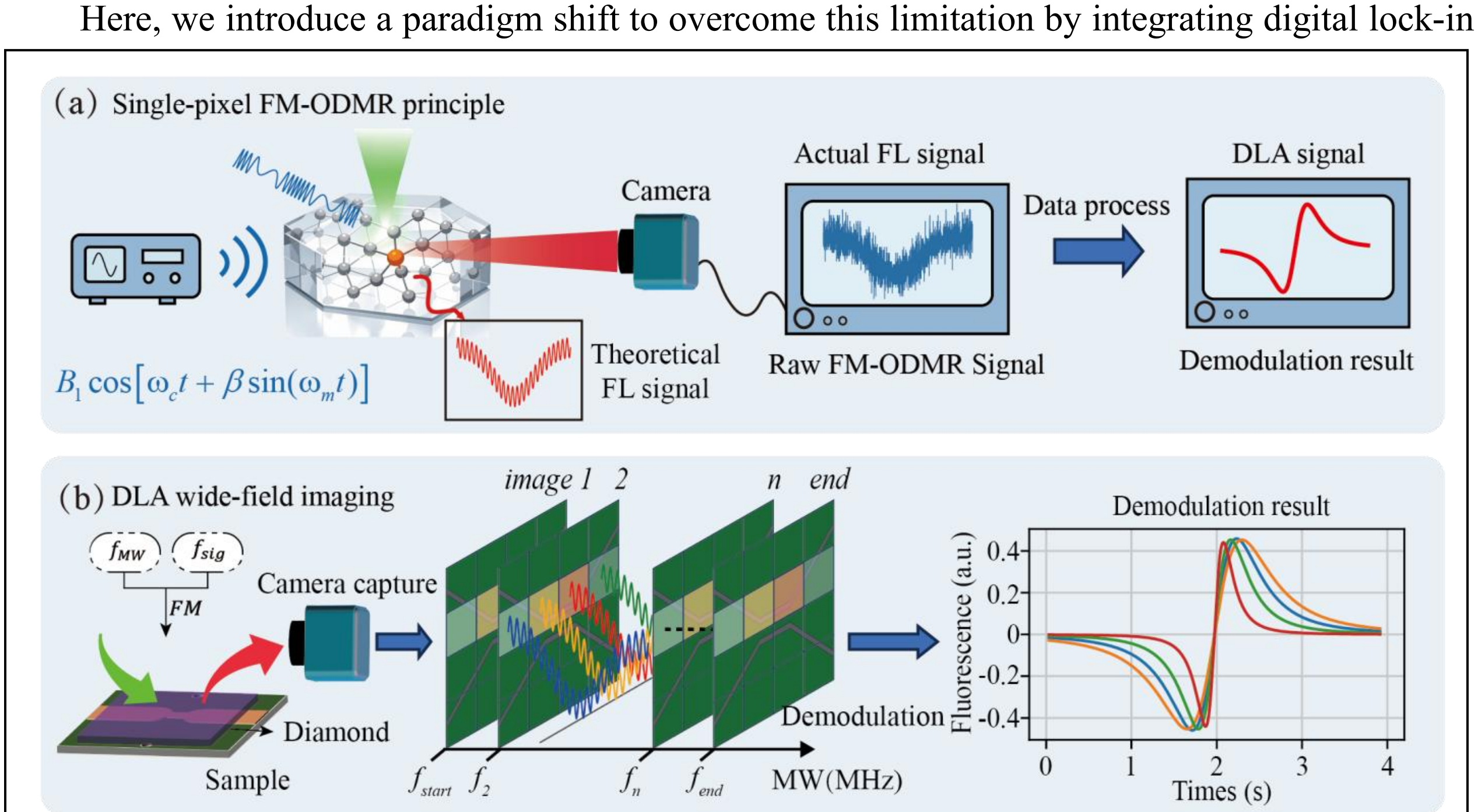

**Fig. 1. Experimental principles.** (a) In the lock-in imaging system, a modulated sinusoidal signal is added to the frequency-swept MW via FM modulation. Under this condition, the fluorescence (modulation) spectrum exhibits a resonance synchronized with the modulation signal. Consequently, the signal acquired by the camera contains the FM modulation information. By employing DLA, this signal can be precisely extracted. (b) Principle Diagram of DLA Imaging. Synchronized fluorescence acquisition under FM-modulated mw excitation. Pixel-wise demodulation of the camera-acquired data. The demodulation results from the four example pixels in the figure demonstrate that the FWHM of the fitting curve in the time domain increases with the magnetic field strength. Imaging is carried out by utilizing the differences in FWHM.

## Experimental Principle and Setup

### 2.1 Experimental principle

NV center are point defects in the diamond lattice, consisting of a substitutional nitrogen atom and an adjacent vacancy. As shown in Fig. 16(b), the core of their physical principle lies in their unique electronic spin triplet ground state ($^3A_2$)). In a zero magnetic field, there is a zero-field splitting of approximately 2.87 GHz between the $m_s = 0$ and $m_s = \pm 1$ energy level[21]. This property enables the precise measurement of MW fields.

When pumped with a 532 nm green laser, electrons are excited to the excited state ( $^3E$ ) and then return to the ground state through two pathways: one is the direct radiative transition, where electrons returning from ( $^3E$ ) to ( $^3A_2$) emit red fluorescence in the range of 637–800 nm; the other is the non-radiative intersystem crossing (ISC) pathway, which is a key characteristic of NV centers. Electrons in the $m_s = \pm 1$ state have a higher probability of undergoing non-radiative transitions back to the $m_s = 0$ ground state via metastable states ($^1A_1$ and $^1E$ ), resulting in a significant reduction in their fluorescence intensity. This spin-dependent fluorescence property allows us to accurately detect changes in the ambient magnetic and MW fields through optical means (ODMR) and convert them into observable fluorescence contrast.

As illustrated in Fig. S16(a), the theoretical ODMR spectrum under conventional frequency-swept MW excitation exhibits an ideal Gaussian lineshape. However, during practical imaging acquisition, the measurement is severely corrupted by multiple noise sources, including laser intensity fluctuations, environmental perturbations, MW source noise, and intrinsic camera readout noise. Consequently, the raw ODMR signal acquired from a single pixel within a single sweep cycle suffers from a significantly degraded SNR, leading to inaccurate spectral fitting. To address this limitation, our DLA imaging system applies a sinusoidal modulation to the frequency-swept MW carrier via frequency modulation (FM) — following Eq.(1). Under this scheme, the detected fluorescence intensity is synchronously modulated in resonance with the modulation signal. As shown in Fig.1(b), the single-pixel, single-cycle ODMR data captured by the camera inherently encodes this modulation frequency, enabling the DLA to precisely extract the signal while effectively suppressing background noise. The principle of our DLA technology is as follows:

Frequency Modulation (FM) of the MW Source:

$$S_{\text{FM}}(t) = A_c \cos\left[2\pi f_c t + \beta \sin(2\pi f_m t)\right] \tag{1}$$

$S_{FM}(t)$ represents the frequency-modulated signal, where $A_c$ denotes the carrier amplitude. The cosine function cos() describes the periodic oscillation characteristic of the signal. $f_c$ is the carrier frequency, which corresponds to the frequency set by the MW source—specifically, the sweep frequency used in the experiment. The following Eq.(7) provides two key parameters closely related to the expression above: the maximum frequency deviation and the modulation index.

$$\begin{aligned} \Delta f &= k_f A_m \\ \beta &= \frac{\Delta f}{f_m} \end{aligned} \tag{2}$$

$A_m$ denotes the amplitude of the sinusoidal modulating signal, and $f_m$ represents the modulating frequency, which serves as the reference frequency for the digital lock-in amplifier. $\Delta f$ is the peak frequency deviation, and $\beta$ is the dimensionless modulation index.

The demodulation formula for the DLA technique is given by:

$$\begin{aligned} I(t) &= s(t) \cdot \sin(2\pi f_m t) \\ Q(t) &= s(t) \cdot \cos(2\pi f_m t) \end{aligned} \tag{3}$$

$s$(t) represent the collected modulated fluorescence signal. It contains a weak component synchronized with the microwave frequency modulation at frequency $f_m$, expressed as $A\cos(2\pi f_m t + \varphi)$, along with superimposed noise $n$(t):

$$s(t) = A\cos(2\pi f_m t + \varphi) + n(t). \tag{4}$$

Orthogonal Mixing – This is the core of phase-locked detection. The signal $s$(t) is multiplied by a reference cosine wave $\cos(2\pi f_m t)$ in the Q-channel (the I-channel is processed analogously):

$$Q(t) = \left[A\cos(2\pi f_m t + \varphi)\right] \cdot \cos(2\pi f_m t) + \cdots. \tag{5}$$

Using the trigonometric identity $\cos\alpha\cos\beta = \frac{1}{2}[\cos(\alpha-\beta) + \cos(\alpha+\beta)]$, the expression becomes:

$$Q(t) = \tfrac{1}{2} A\left[\cos\varphi + \cos(4\pi f_m t + \varphi)\right] + \cdots. \tag{6}$$

This step decomposes the signal into a DC component $\left(\frac{1}{2}A\cos\varphi\right)$ and a double-frequency component $\left(\frac{1}{2}A\cos(4\pi f_m t + \varphi)\right)$. Low-Pass Filtering – As described by Eq.(12), the low-pass filter then "removes the high-frequency part while retaining the low-frequency part."

$$\begin{aligned} I_{\text{filtered}}(t) &= \text{LPF}\{I(t)\} \\ Q_{\text{filtered}}(t) &= \text{LPF}\{Q(t)\} \end{aligned} \tag{7}$$

The low-pass filter removes the double-frequency component (the $2f_m$term) and high-frequency noise $n$(t), while retaining the DC components that carry the amplitude and phase information of the original signal. Consequently, after reaching steady state, the filtered signals $I_{\text{filtered}}$ and $Q_{\text{filtered}}$ are respectively given by:

$$I_{\text{filtered}} = -\frac{1}{2} A \sin\varphi, \qquad Q_{\text{filtered}} = \frac{1}{2} A \cos\varphi. \tag{8}$$

By processing the camera-acquired signals using the equations above, we obtain the final demodulated profile displayed in Fig.1(b). In the experiment, the MW field is calculated by fitting $Q_{\text{filtered}}(t)$ using Eq.(9) to obtain the parameter $\omega$.

$$y(x) = -\frac{4A}{\omega^3 \sqrt{\pi/2}} \cdot (x - x_c) \cdot \exp\left[-2\left(\frac{x - x_c}{\omega}\right)^2\right] \tag{9}$$

This formula represents the first derivative $dG(x)/\text{d}x$ of a Gaussian function $G(x)$ with a specific form (the full derivation is provided in Supplementary Material S2). Here, $x_c$ is the center position of the above expression and also corresponds to the zero-crossing point of its derivative. $\omega$ is the FWHM. $A$ is the area parameter – it is not the peak height of the derivative curve, but rather the total integrated area of the parent Gaussian function $G(x)$. The parameter $\omega$ denotes the width parameter, defined as the distance between its maximum and minimum values, and serves as a key variable for calculating the microwave field strength.

The MW field strength $B_{\text{MW}}$ is given by (detailed derivation can be found in Supplementary Section S3)[14,21,25]:

$$B_{\text{MW}} \approx \frac{\sqrt{\left(\omega_{\text{FWHM}} - \frac{1}{\pi T_2^*}\right)^2 - \left(\frac{1}{T_2} + \frac{\Gamma_p}{2}\right)^2 / \pi^2}}{2\pi\gamma_{\text{NV}} \cdot \sqrt{4\left(1/T_2 + \Gamma_p/2\right)/\left(1/T_1 + \Gamma_p\right)}} \tag{10}$$

Here, $T_1$ denotes the lattice (spin–lattice) relaxation time, $T_2$ the spin coherence time, and $T_2^*$ the inhomogeneous dephasing time. These parameters $T_1$, $T_2$, and $T_2^*$ are characteristic constants determined by the specific NV centers sample and its local environment. In addition, $\Gamma_p$ represents the optical pumping rate set by the laser excitation, and $\gamma_{\text{NV}}$ is the gyromagnetic ratio of the NV-center electron spin, with a value of $\gamma_{\text{NV}} \approx 28\ \text{MHz/mT}$. From the expression above, it can be observed that the $\omega$ of the resonance line scales approximately linearly with the MW magnetic field strength, i.e., the linewidth broadens as the MW field increases (under weak-driving conditions).

Fig.1(b) illustrates the working principle of DLA imaging. By fitting the spectral curve extracted from the acquired camera frames using Eq.(10), the corresponding linewidth ω is directly determined. Spatial imaging is subsequently achieved by mapping the pixel-wise distribution of ω. This procedure enables the two-dimensional visualization of the microwave magnetic field distribution over the sample surface.

## 2.2 Construction and System Characterization of the DLA Imaging Experimental Platform

To achieve the two-dimensional visualization and characterization of the microwave magnetic field distribution on the surface of microstrip circuits, we constructed a DLA imaging system based on a NV centers microscope.

As shown in Fig. 2(a), the light source of the digital lock-in imaging system is a 532-nm laser (cnilaser, MLL-V-532). The laser beam first passes through an optical isolator and polarization control optics. It is then directed through an acousto-optic modulator (used to generate pulsed laser control) and subsequently guided by a series of mirrors. The beam is expanded by a beam expander (JCOPTIX, OSE05-532) and homogenized to form a suitably sized spot. Afterwards, the beam is reflected by a dichroic mirror and focused by an objective lens onto the surface of a diamond sample that is placed in close contact with the device under test, as illustrated in Fig. 2(b).

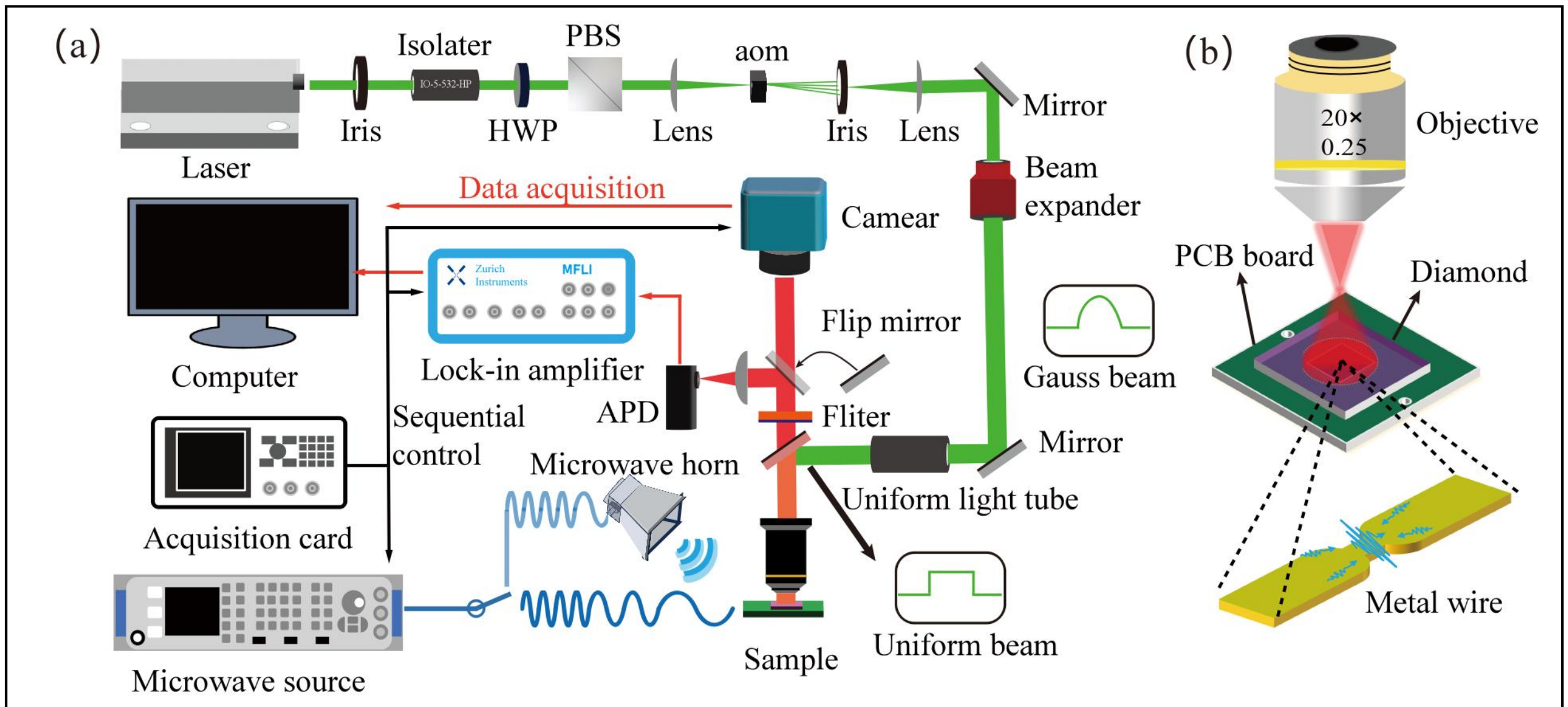

**Fig. 2. Experimental platform.** (a) Optical path diagram of the imaging system. (b) Schematic diagram showing a diamond sample positioned above the region of interest, where it is excited by green light and the resulting red photoluminescence is collected by an objective lens.

When the NV centers in the diamond are excited, they emit red photoluminescence (PL). This PL carries the modulated microwave information and is collected by a high-speed camera (MARS-251-1938X2M/C-NF, full-frame maximum frame-rate limit: 1937 fps). The signals are then transferred to a computer for digital lock-in amplification analysis and processing, as shown in Fig. 3. In addition, a flip mirror is incorporated to direct a portion of the PL to a balanced photodetector (LBTEK, PDBS1A-DC). The electrical signal from the detector is processed by a lock-in amplifier (MFLI, 5 MHz) and the resulting data is also sent to the computer.

In the measurement system, a reference frequency is modulated onto the frequency-swept microwave generated by the source. The microwave signal is produced by an analog signal generator (N5181B MXG), amplified by a power amplifier (Talent Microwave, TLPA1G6G-43-43), and then either directly fed into the device under test or radiated onto the sample surface via a horn antenna, as depicted in Fig. 3.

Finally, based on the principle illustrated in Fig. 2, we constructed a digital lock-in imaging system utilizing NV centers. The system achieved a field of view of 2048 × 1216 pixels with a spatial resolution of 1.6 μm (the resolution analysis is provided in Supplementary Material S1). After the system was established, we innovatively integrated the theory and technique of lock-in amplification with wide-field imaging, enabling the visualization of microwave fields from microstrip structures. The specific procedure for two-dimensional characterization is outlined in Fig. 1(b).

Additionally, we propose two distinct measurement schemes:

(1) Inline measurement – when the frequency of the device under test (DUT) falls within the detection bandwidth of the system, the microwave magnetic field imaging is performed by directly feeding the MW signal into the DUT via a transmission line.

(2) Radiative measurement – when the operating frequency of the DUT lies outside the system's detection range, the MW field is delivered through a horn antenna that irradiates the sample surface, as illustrated in Fig. 3[26,27].

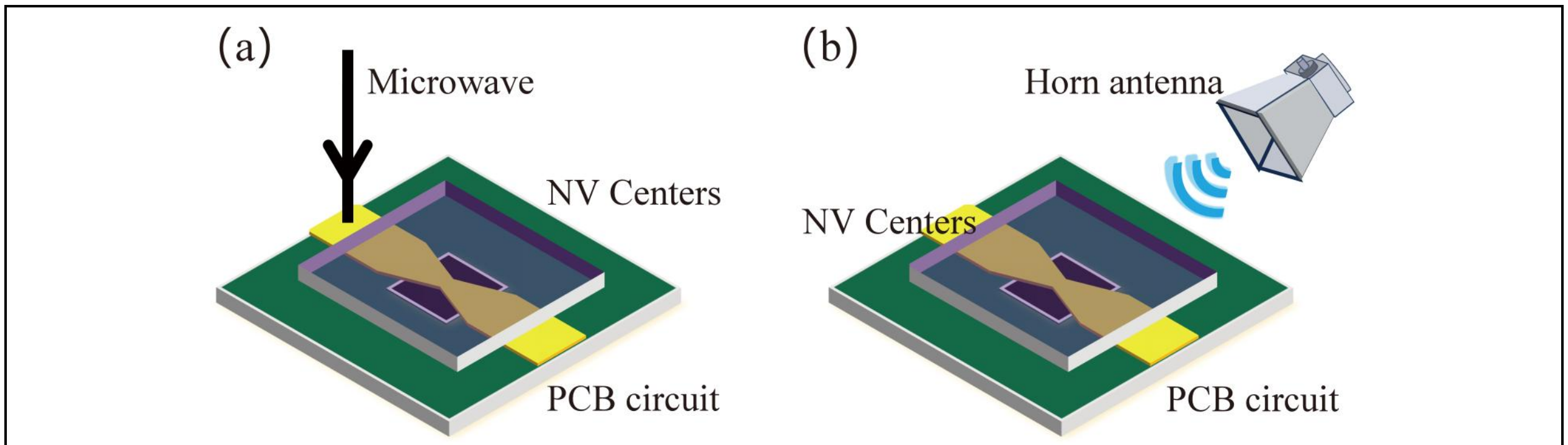

**Fig. 3. Two methods of microwave supply.** (a) Microwave supply via direct cable connection. (b) Microwave supply via a horn antenna.

## RESULTS

### 3.1 Experimental at 2.87 GHz

In the experiment, our system acquired the demodulation spectra corresponding to the MW power range from 8 dBm to 23 dBm. The linewidth ω was obtained through fitting, and the measured results were compiled and plotted in Fig. 4. Since the output power of the commercial microwave source is conventionally calibrated in logarithmic units (dBm), we first converted the power from dBm to linear units (W) using Eq. (11) to establish a direct correlation between the experimental parameters and the theoretical model, thereby facilitating subsequent physical analysis based on power and the inversion of magnetic field strength.

$$P = 10^{\left(\frac{P_{\mathrm{dBm}}}{10}\right)} \times 10^{-3} \tag{11}$$

Subsequently, the power in watts was converted to the corresponding magnetic field strength using Eq. (12):

$$B_{\mathrm{RMS}} = \frac{\sqrt{2\mu_0 P}}{v_p w h} \tag{12}$$

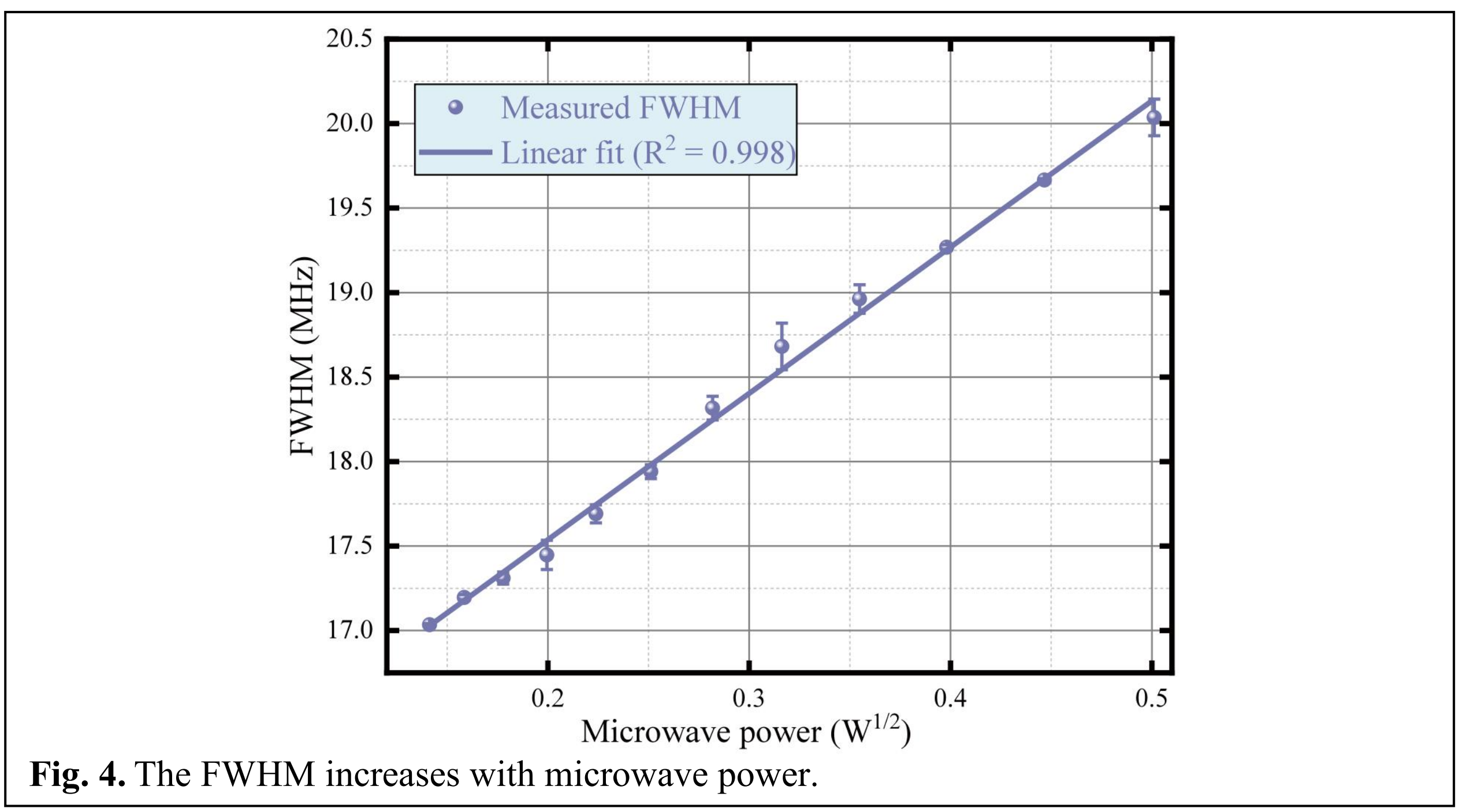

**Fig. 4.** The FWHM increases with microwave power.

According to the theoretical derivation in Eq. (12), the microwave magnetic field strength *B* is proportional to the square root of the microwave power $\sqrt{P}$. Meanwhile, combined with the model prediction from Eq. (10), the linewidth ω exhibits an approximately linear relationship with *P*. This trend is in excellent agreement with the experimentally measured variation of the linewidth with power, as shown in Fig. 4.

Building upon the test results corresponding to different microwave power levels presented earlier, this study selects four representative sets of experimental data within the 8 dBm to 23 dBm range(experimental procedures can be found in the "Experimental Methods" section). A normalized imaging algorithm and data optimization are applied to visually illustrate the imaging variations under different driving conditions, as specifically shown in Fig. 5. In the figure, the coordinate axes represent pixel positions, where one pixel corresponds to an actual distance of 1.6 μm.

As previously mentioned, if the resonance frequency of the device under test deviates from 2.87 GHz, or if in situ microwave-coupled measurements cannot be performed under operational conditions, the spatially radiated scheme using a horn antenna shown in Fig. 3(b) can be adopted. The underlying principle of this scheme is that metallic structures absorb the spatially propagating microwave signal, thereby modulating the spin energy levels of the NV centers and enabling contactless microwave driving. Compared to the local MW excitation method where the signal is directly coupled into the sample, the horn-antenna radiation approach exhibits a relatively lower coupling efficiency. Hence, in the experiment, the output power of the horn antenna was set to 23 dBm to compensate for the signal loss, and the corresponding test results are presented in Fig. 6.

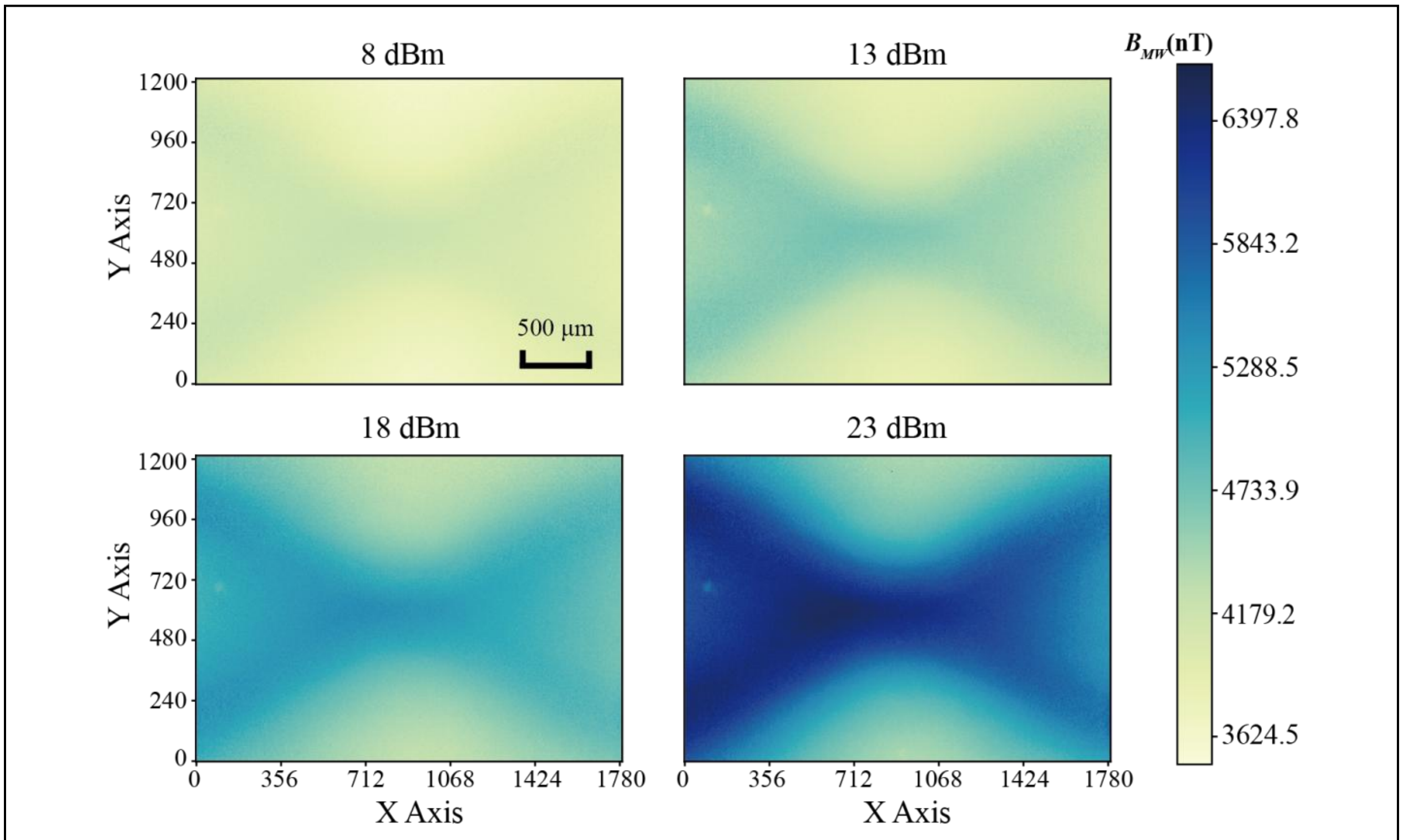

**Fig. 5.** Imaging results of the tested microstrip antenna at power levels ranging from 8 to 23 dBm(microwave is input into the microstrip antenna through the cable.).

As shown in Fig. 6, although the magnetic field strength effectively excited by the system is significantly reduced when the microwave is delivered via the spatially radiating horn antenna, this approach can still effectively delineate the boundaries of metallic structures by leveraging the magnetic-field-dependent response of NV centers, thereby achieving a field-visualization mapping of the metallic contours.

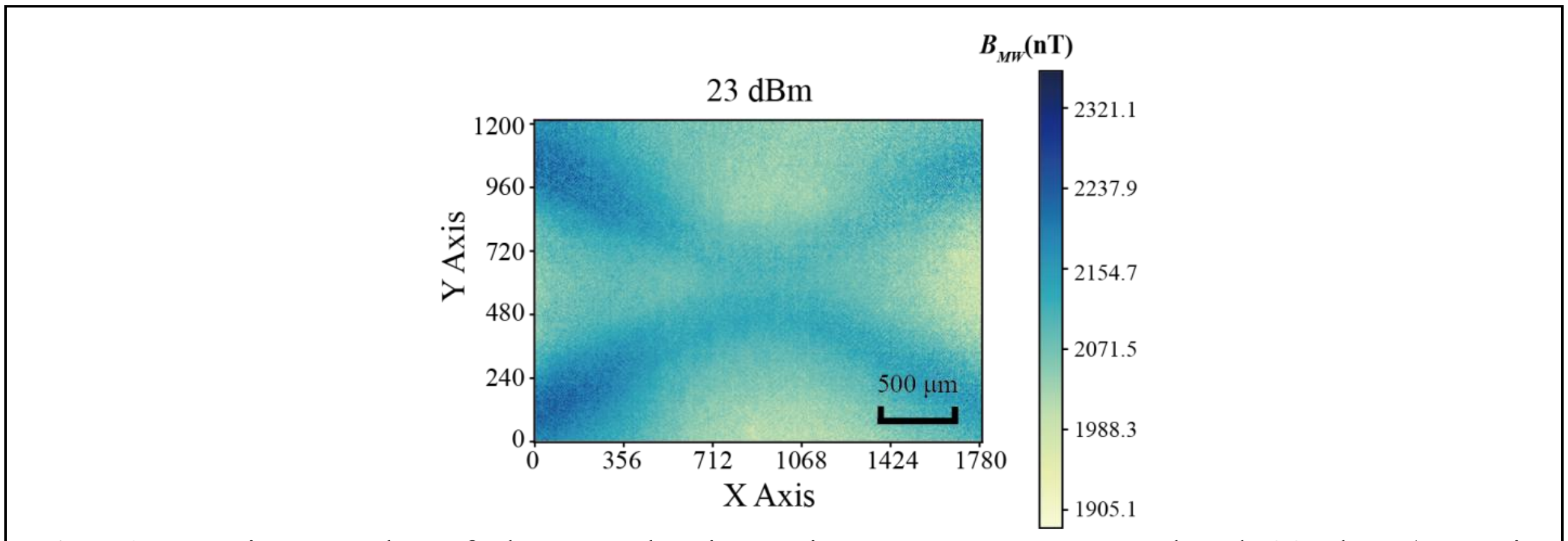

**Fig. 6.** Imaging results of the tested microstrip antenna at power level 23 dBm(MW is transmitted through the horn antenna and onto the microstrip antenna.).

Having completed the microwave magnetic field imaging experiments and the corresponding analysis under different working conditions, this work further validates the accuracy and effectiveness of the experimental measurements. By employing COMSOL Multiphysics software, we modeled and simulated the MW magnetic field distribution characteristics of the sample. The simulated and experimentally measured results were then normalized and compared, as shown in the overall comparison presented in Fig. 7(a). Building on this, a one-dimensional line profile extracted at the lateral coordinate of 120 pixels was used for a direct comparison between the simulation and experimental data, as depicted in Fig. 7(b), where the red curve represents the simulated data and the yellow curve corresponds to the measured results. It can be observed that the experimental and simulated curves exhibit a high degree of agreement in their overall trends and distribution patterns, thereby effectively verifying the reliability and accuracy of the experimentally obtained microwave magnetic field imaging results. This demonstrates that the proposed imaging and measurement scheme offers good feasibility and stability.

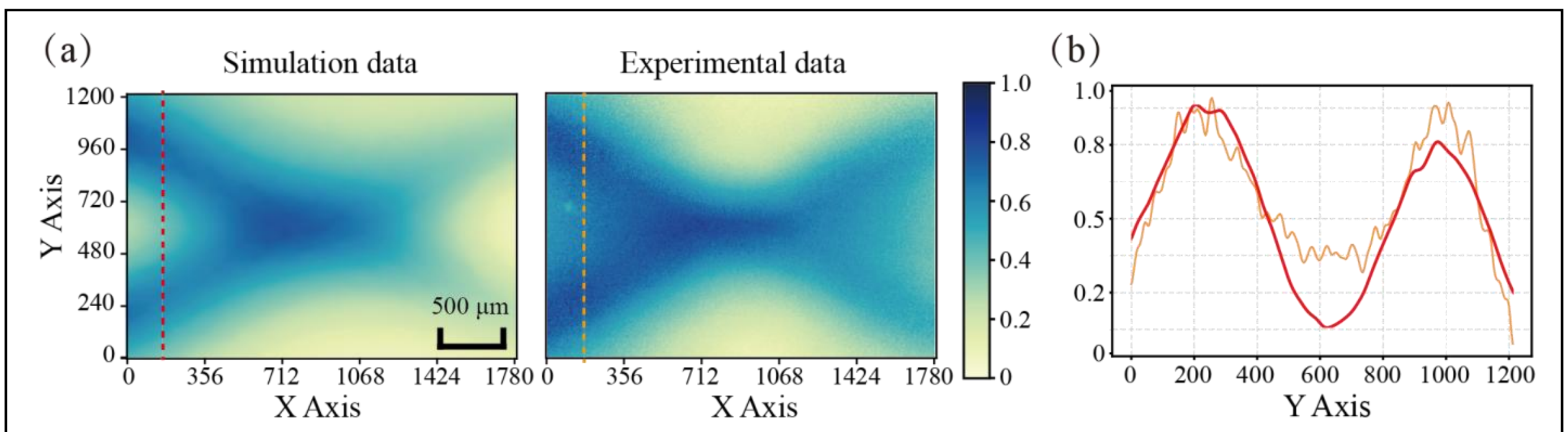

**Fig. 7. Simulation vs. Experimental Results.** (a) Experimental result and COMSOL simulation result (at 23 dBm). (b) Comparison of line profiles (red/orange) at the same position from simulation and experiment.

### 3.2 Experimental at 2.73 GHz and 3.0 GHz

The ensemble of NV centers in diamond possesses four inherent spatial orientations. Under an applied static magnetic field, the electron spin energy levels corresponding to each NV axis undergo Zeeman splitting[28]. Therefore, with the proper application of a directional bias magnetic field, the ODMR spectrum can split into eight distinct resonance dips[29,30]. Based on this physical mechanism, the effective detection bandwidth for microwave excitation can be significantly broadened by controlling the magnitude and direction of the external static field[31]. In this experiment, a small permanent magnet was employed to apply a controllable bias magnetic field to the NV center ensemble, achieving controlled splitting of the resonance dips. The resulting split ODMR spectral features are shown in Fig. 8(a). Furthermore, utilizing a novel microwave radiating structure, microwave field imaging tests were conducted at resonance frequencies of 2.73 GHz and 3.0 GHz, located at the two ends of the spectrum. The microwave output power was uniformly set to 23 dBm for these experiments. The final dual-frequency microwave field imaging results are presented in Fig. 8(b).

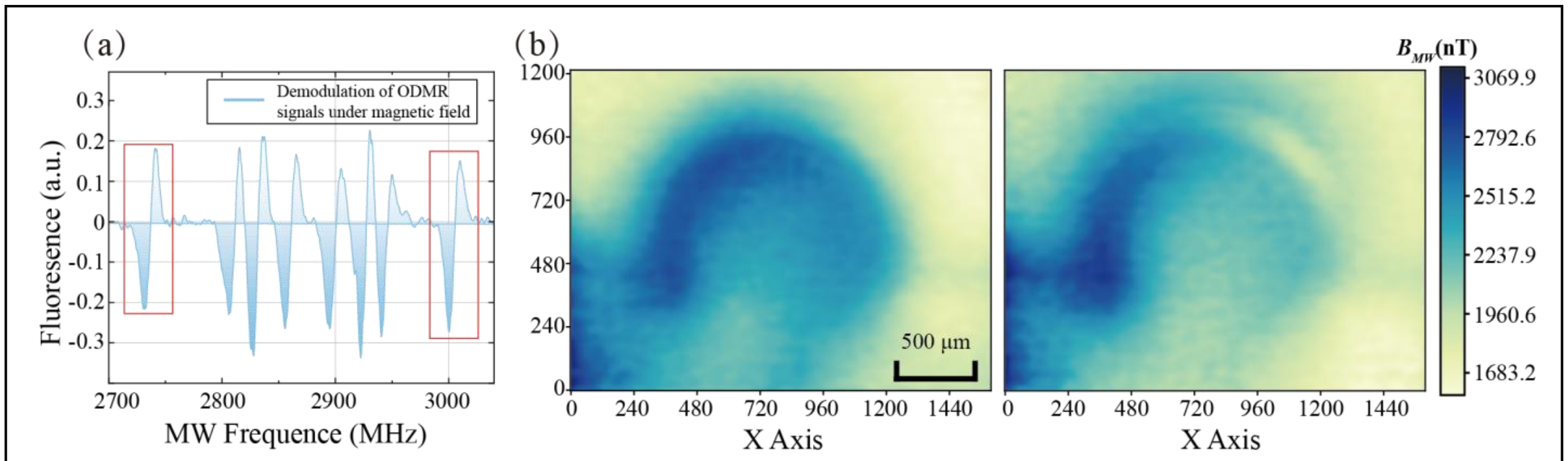

**Fig. 8. Bandwidth testing.** (a) Applies a bias magnetic field to the diamond, splitting the original NV centers peak into eight. The modulation and demodulation results for the eight peaks. (b) Shows the microwave field imaging results corresponding to the two outermost peaks.

In Fig. 8(b), the left panel shows the microwave field imaging result corresponding to 2.73 GHz, while the right panel displays the test result for 3.0 GHz. Comparing the two sets of imaging characteristics, it is clearly evident that the radiative coupling efficiency and field excitation performance of the microwave radiating structure are significantly better at the 2.73 GHz frequency band than at 3.0 GHz.

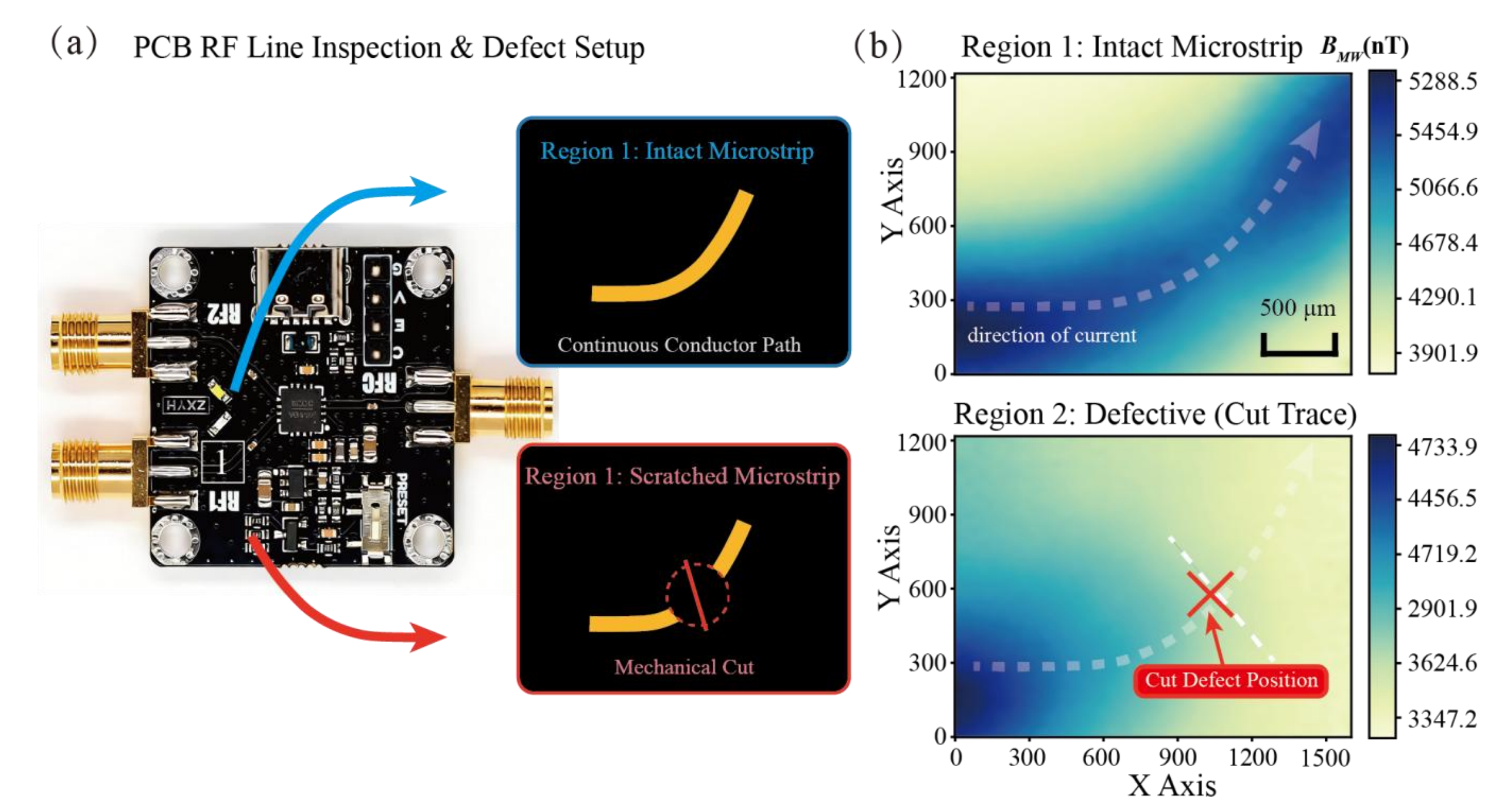

**Fig. 9. Application testing.** (a) An RF switch microstrip circuit serves as the test device, with one areas (blue: normal, red: anomalous) defined for evaluation. (b) Test results of the normal and anomalous regions.

### 3.3 Applied Experiment

To validate the practical engineering value of this magnetic field imaging technique, this paper selects a commercial RF switch as the device under test for open-circuit fault detection and imaging experiments. The operational frequency range of this RF switch spans from 9 kHz to 6 GHz, fully covering the microwave frequency bands used in this study, thus ensuring excellent compatibility. For the experiment, on the RF switch were selected and defined as Region 1, respectively, with the imaging sampling locations shown in Fig. 9(a). First, we measured the MW magnetic field in Region 1 under its fault-free operating condition. Subsequently, we manually scratched the surface of Region 1 using a tool to create a short-circuit fault. Finally, we performed MW magnetic field imaging on Region 1. The corresponding imaging results are presented in Fig. 9(b). Region 1, subjected to the scratch, shows no discernible magnetic field signal, indicating a disrupted magnetic flux path, which stands in clear contrast to the intact Region 1. These results demonstrate that the NV-center-based microwave magnetic field imaging technique can visually identify open-circuit defects and effectively enable the visual diagnosis of line faults in RF devices, highlighting its significant potential for practical engineering applications.

## DISCUSSION

Based on the data obtained from previous experiments, combined with the power spectral density. The sensitivity of the digital lock-in imaging technique based on NV centers can be expressed by the following formula[31–34]:

$$\eta = \frac{\sigma_B}{\sqrt{\text{ENBW}}} = \frac{\sqrt{\sigma_{\text{SN}}^2 + \sigma_L^2 + \sigma_E^2}}{k\gamma_e\sqrt{\text{ENBW}}} = \frac{\text{ASD}}{\max\left(\left|\frac{d(a.u.)}{df}\right|\right)\gamma_e} \tag{13}$$

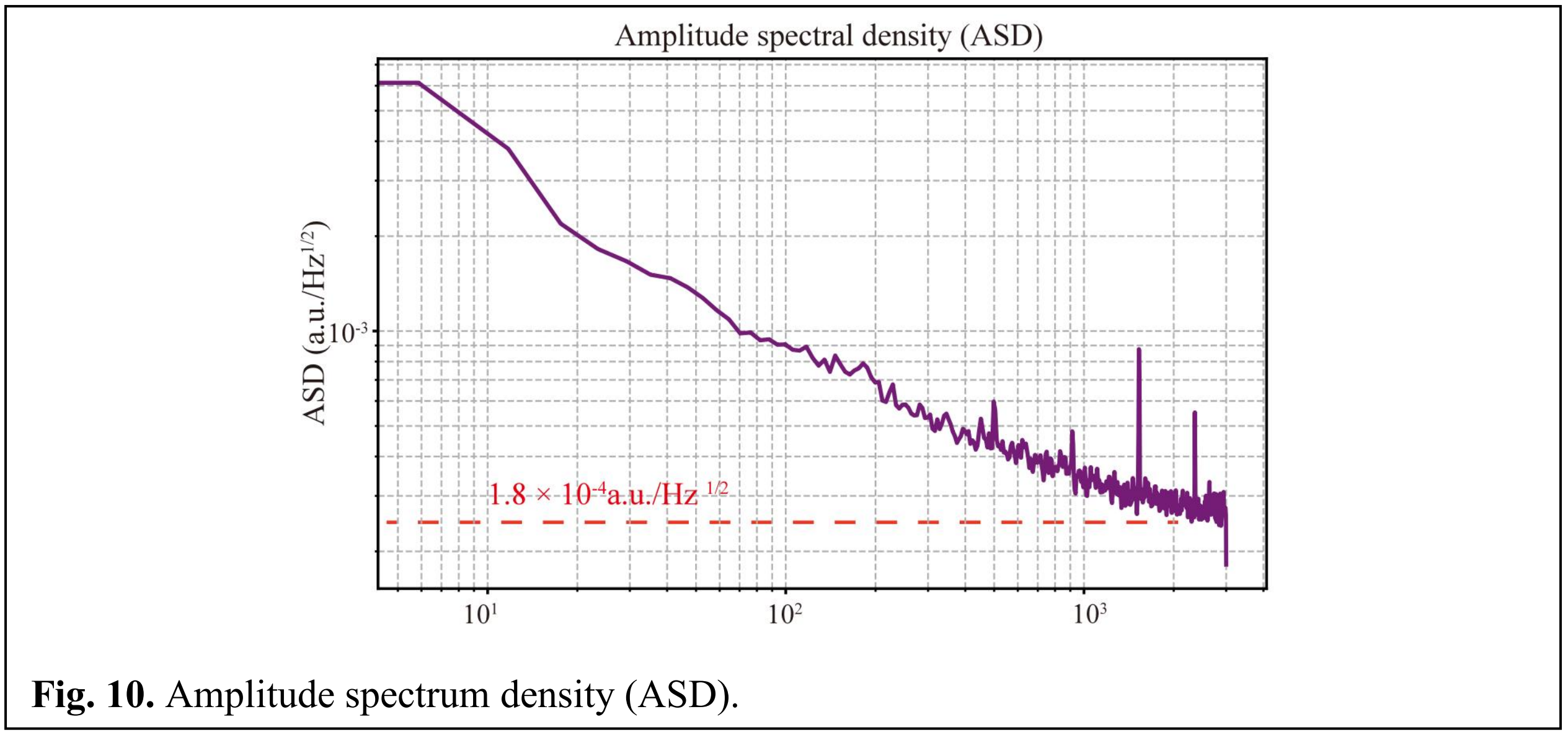

**Fig. 10.** Amplitude spectrum density (ASD).

In the expression, $\sigma_B$ denotes the system magnetic noise, ENBW is the equivalent noise bandwidth of the system, $\sigma_{\text{SN}}$ represents the photon shot noise, $\sigma_L$ stands for the laser noise, $\sigma_E$ corresponds to the system electronic noise, $k$ is the maximum slope of the demodulated signal, and $\gamma_e$ refers to the electron gyromagnetic ratio of the NV center, given as $2.8 \times 10^{-5}$ MHz/nT. ASD (amplitude spectral density) describes the amplitude spectral density of the demodulated signal, which characterizes the system noise per unit frequency. The amplitude spectral density of our system is illustrated in Fig. 10, where the maximum slope, $\max\left(\left|\frac{d(a.u.)}{df}\right|\right)$, is measured to be $0.051\ (a.u.)/\mathrm{MHz}$. Based on the calculation, the sensitivity is determined to be $126\ \mathrm{nT/Hz^{1/2}}$.

This paper proposes and implements a novel wide-field microwave magnetic field imaging method based on diamond NV centers combined with digital lock-in amplification. By applying frequency modulation to the microwave source and demodulating the pixel-wise fluorescence signals collected by the camera using a custom-developed digital lock-in program, weak signals synchronous in both frequency and phase with the microwave field response are efficiently extracted in the presence of complex background noise. The system achieves a magnetic field detection sensitivity of $126\ \mathrm{nT/Hz^{1/2}}$, a spatial resolution of $1.6\mathrm{\mu m}$, and a full-field imaging time on the order of seconds. With a field of view of $2048\ \times 1216$ pixels, two experimental schemes were designed: first, online imaging of a microwave-activated microstrip line via its input feed line; second, non-contact radiative imaging of a passive microstrip line using a horn antenna. These approaches successfully enable two-dimensional visualization of the microwave field distribution over microstrip circuit surfaces. Moreover, artificially introduced open-circuit defects were precisely located and diagnosed, demonstrating the method's capability for fault analysis. This technique provides a high-sensitivity, high-resolution, and rapid imaging tool for non-destructive testing and failure analysis of high-performance radio-frequency devices[34].

## MATERIALS AND METHODS

### 5.1 Diamond preparation

The diamond used in the experiments was sourced from Element Six, with dimensions of 3 mm × 3 mm × 0.5 mm. We measured relaxation times of $T_1 \approx 7.09$ ms, $T_2 \approx 9.89$ μs, and $T_2^* \approx 644.89$ ns, optical pumping rate $\Gamma_p = 205.6$ Hz (its detailed parameters are provided in Supplementary Material S4) .

### 5.2 Experimental Methods

In Fig. 1(b), the NV centers in the diamond placed on the sample surface are first excited by a 532 nm laser, and the emitted photoluminescence (PL) is captured by a high-speed camera and saved as a video file in AVI format. Subsequently, the recorded video is processed using Python: each frame is converted into a grayscale matrix indexed by pixel coordinates. By maintaining synchronized data acquisition, the PL signal at each identical pixel index across all frames is extracted to generate a time-dependent PL curve for that specific location. This time-domain curve can also be transformed into the frequency domain, which contains information about the magnetic field strength. By fitting this curve using Eq. (9), the corresponding linewidth ω is directly obtained.

In the experiment centered at 2.87 GHz, the MW source was configured with the following parameters: the frequency sweep range was set from 2.83 GHz to 2.91 GHz, with a step size of 200 kHz, resulting in a total of 400 frequency points. The trigger frequency of the microwave source was 100 Hz. Meanwhile, according to Eq. (2), the modulation frequency $f_m$ and amplitude $A_m$ were set to 100 Hz and 1 V, respectively, and the modulation frequency deviation $\Delta f$ was set to 8 MHz. For data acquisition, the sampling rate of the high-speed camera was set to 600 Hz, and a complete single imaging cycle took $4\,s$ (400 points / 100 Hz). A comparison with other works[35–44] is shown in Fig.11. Our imaging time is superior to that of other works, and we have maintained an excellent spatial resolution in a large field of view.

Analysis indicates that the main bottleneck in imaging speed lies in the microwave sweep time, which is jointly determined by the sweep range, the number of frequency points, and the trigger frequency. Therefore, by reducing the sweep range, decreasing the number of frequency points, and increasing the trigger frequency, the imaging speed can be significantly improved, though this also imposes higher demands on the sampling rate of the high-speed camera.

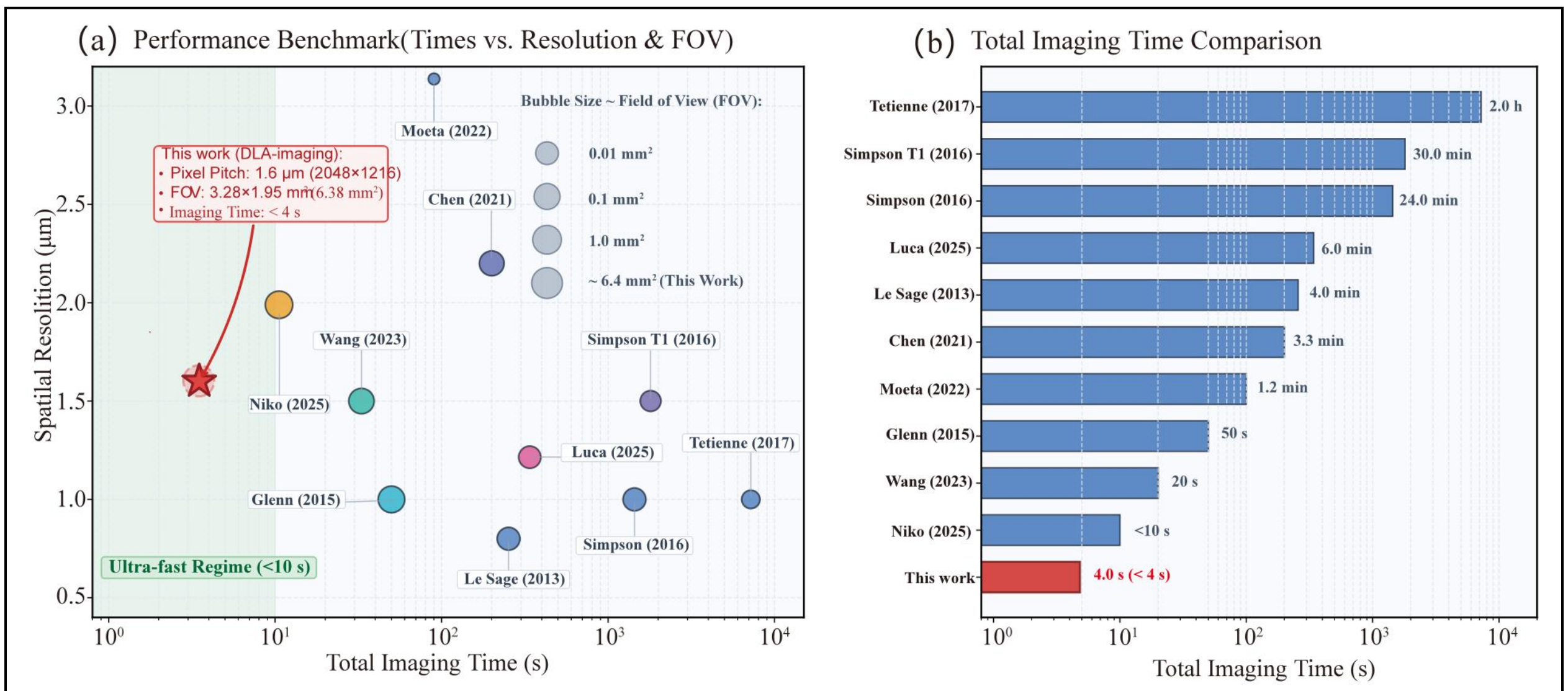

**Fig. 11.** Benchmark of wide-field NV quantum imaging performance compared with representative literature.(a) Trade-off map of spatial resolution versus total acquisition time across representative wide-field NV sensing modalities. The bubble area scales logarithmically with the optical field of view (FOV). The light-green shaded area indicates the ultra-fast imaging regime (< 10 s). Our work (red star) achieves sub-4-second full-spectrum mapping over a millimeter-scale FOV (6.4 $mm^2$).(b) Acquisition time ranking on a logarithmic scale, showing orders-of-magnitude acceleration achieved by our digital lock-in amplifier (DLA) demodulation scheme.

Hence, the 4 S imaging time in this work serves only as a reference; it can indeed be made faster. For the experiments with center frequencies of 2.73 GHz and 3.0 GHz, the microwave sweep ranges were adjusted to 2.69 GHz–2.77 GHz and 2.96 GHz–3.04 GHz, respectively, while all other experimental parameters remained consistent with those used at 2.87 GHz.

## ACKNOWLEDGMENTS

**Author contributions:** Z.F. designed the experimental protocol. Z.F. and Y.L. performed the experiments; Y.H., H.L., Z.W., and W.H. analyzed the experimental results. Z.F. wrote the manuscript. Z.F., Y.L., B.Z. supervised the project, and provided ideas and necessary technical support. Z.F. and Y.L. made the same contribution to this work. All authors reviewed the manuscript.

**Funding:** This study was supported by NSF of China (Grant No.T2522007, No.12441502, No.12374462)

**Competing interests:** The authors declare no competing financial interest.

## DATA AVAILABILITY

The data supporting the findings of this study are available from the corresponding authors upon reasonable request.

# REFERENCES

1. Cengiz, A., Bilim, M. & Kabalci, Y. The Next Generation in Communication Technology: Roadmap to 6G. in *2024 6th Global Power, Energy and Communication Conference (GPECOM)* 724–729 (IEEE, Budapest, Hungary, 2024). doi:10.1109/GPECOM61896.2024.10582586.
2. Alibakhshikenari, M. *et al.* Study on improvement of the performance parameters of a novel 0.41–0.47 THz on-chip antenna based on metasurface concept realized on 50 μm GaAs-layer. *Sci Rep* **10**, 11034 (2020).
3. Alibakhshikenari, M. *et al.* Author Correction: High-Gain Metasurface in Polyimide On-Chip Antenna Based on CRLH-TL for Sub-Terahertz Integrated Circuits. *Sci Rep* **10**, 13650 (2020).
4. Yoonjin Won, Jungwan Cho, Agonafer, D., Asheghi, M. & Goodson, K. E. Fundamental Cooling Limits for High Power Density Gallium Nitride Electronics. *IEEE Trans. Compon., Packag. Manufact. Technol.* **5**, 737–744 (2015).
5. Park, S.-H., Choi, S., Song, D.-G. & Jhang, K.-Y. Microstructural Characterization of Additively Manufactured Metal Components Using Linear and Nonlinear Ultrasonic Techniques. *Materials* **15**, 3876 (2022).
6. Wu, R., Zhang, H., Yang, R., Chen, W. & Chen, G. Nondestructive Testing for Corrosion Evaluation of Metal under Coating. *Journal of Sensors* **2021**, 6640406 (2021).
7. Cao, Y., Chen, K. & Ruan, C. A Microwave Metamaterial-inspired Sensor for Nondestructive Evaluation of Dielectric Substrates. in *2021 International Conference on Microwave and Millimeter Wave Technology (ICMMT)* 1–3 (IEEE, Nanjing, China, 2021). doi:10.1109/ICMMT52847.2021.9617917.
8. Liu, H., Tan, C., Zhao, S. & Dong, F. Nonlinear Ultrasonic Transmissive Tomography for Low-Contrast Biphasic Medium Imaging Using Continuous-Wave Excitation. *IEEE Trans. Ind. Electron.* **67**, 8878–8888 (2020).
9. Shao, W. *et al.* Simultaneous Measurement of Electric and Magnetic Fields With a Dual Probe for Efficient Near-Field Scanning. *IEEE Trans. Antennas Propagat.* **67**, 2859–2864 (2019).
10. Wen, H. F. *et al.* Recognizing Microwave Field Contrast of Invisible Microstrip Defects With High Accuracy by Quantum Wide-Field Microscope. *IEEE Trans. Microwave Theory Techn.* **72**, 5896–5903 (2024).
11. Zhang, H. *et al.* Separation and imaging of mixed signals on microwave chips based on nitrogen-vacancy color center microscopy. *Photon. Res.* **13**, 1200 (2025).
12. Hong, S. *et al.* Nanoscale magnetometry with NV centers in diamond. *MRS Bull.* **38**, 155–161 (2013).
13. Ma, L. *et al.* A Magneto-Optically Co-Enhanced Direct-Readout Diamond NV Fiber-Optic Vector Quantum Magnetometer. *Laser & Photonics Reviews* e03258 (2026) doi:10.1002/lpor.202503258.
14. Grinolds, M. S. *et al.* Nanoscale magnetic imaging of a single electron spin under ambient conditions. *Nature Phys* **9**, 215–219 (2013).
15. López-Morales, G. I., Zajac, J. M., Flick, J., Meriles, C. A. & Dreyer, C. E. Quantum embedding study of strain- and electric-field-induced Stark effects on the NV − center in diamond. *Phys. Rev. B* **110**, 245127 (2024).
16. Kehayias, P. *et al.* Imaging crystal stress in diamond using ensembles of nitrogen-vacancy centers. *Phys. Rev. B* **100**, 174103 (2019).
17. Kucsko, G. *et al.* Nanometre-scale thermometry in a living cell. *Nature* **500**, 54–58 (2013).
18. Neumann, P. *et al.* High-Precision Nanoscale Temperature Sensing Using Single Defects in Diamond. *Nano Lett.* **13**, 2738–2742 (2013).
19. Basso, L. *et al.* Wide-field microwave magnetic field imaging with nitrogen-vacancy centers in diamond. *Journal of Applied Physics* **137**, 124401 (2025).

20. Chang, K., Eichler, A., Rhensius, J., Lorenzelli, L. & Degen, C. L. Nanoscale Imaging of Current Density with a Single-Spin Magnetometer. *Nano Lett.* **17**, 2367–2373 (2017).

21. Li, X. *et al.* A highly integrated portable quantum wide-field microscope using NV centers. *Optics & Laser Technology* **193**, 114213 (2026).

22. Appel, P., Ganzhorn, M., Neu, E. & Maletinsky, P. Nanoscale microwave imaging with a single electron spin in diamond. *New J. Phys.* **17**, 112001 (2015).

23. Moreva, E. *et al.* Magnetic Sensing with Nitrogen-Vacancy Centers Based on Lock-in Detection. in *2020 Conference on Precision Electromagnetic Measurements (CPEM)* 1–2 (IEEE, Denver (Aurora), CO, USA, 2020). doi:10.1109/CPEM49742.2020.9191789.

24. Lei, Y. *et al.* The response optimization for improving the speed of resonance frequency tracking method using nitrogen-vacancy center. *Optics & Laser Technology* **190**, 113260 (2025).

25. Jensen, K., Acosta, V. M., Jarmola, A. & Budker, D. Light narrowing of magnetic resonances in ensembles of nitrogen-vacancy centers in diamond. *Phys. Rev. B* **87**, 014115 (2013).

26. Chen, X.-D. *et al.* Focusing the electromagnetic field to 10−6λ for ultra-high enhancement of field-matter interaction. *Nat Commun* **12**, 6389 (2021).

27. Zhu, Q. *et al.* Identifying subsurface metal microstructure and its materials via quantum wide-field microscope. *Opt. Express* **33**, 51956 (2025).

28. Wang, B. *et al.* Simultaneous detection of multi-channel signals in MHz bandwidth using nitrogen-vacancy centers in a diamond. *Opt. Express* **32**, 3184 (2024).

29. Garsi, M. *et al.* Three-dimensional imaging of integrated-circuit activity using quantum defects in diamond. *Phys. Rev. Applied* **21**, 014055 (2024).

30. Turner, M. J. *et al.* Magnetic Field Fingerprinting of Integrated-Circuit Activity with a Quantum Diamond Microscope. *Phys. Rev. Applied* **14**, 014097 (2020).

31. Zhu, X. *et al.* Joint quantum sensing of vector magnetic field and temperature with nitrogen-vacancy centers in diamond. *Applied Physics Letters* **123**, 244002 (2023).

32. Ahmadi, S., El-Ella, H. A. R., Hansen, J. O. B., Huck, A. & Andersen, U. L. Pump-Enhanced Continuous-Wave Magnetometry Using Nitrogen-Vacancy Ensembles. *Phys. Rev. Applied* **8**, 034001 (2017).

33. Xie, Y. *et al.* A hybrid magnetometer towards femtotesla sensitivity under ambient conditions. *Science Bulletin* **66**, 127–132 (2021).

34. Kim, D. *et al.* A CMOS-integrated quantum sensor based on nitrogen–vacancy centres. *Nat Electron* **2**, 284–289 (2019).

35. Tsukamoto, M. *et al.* Accurate magnetic field imaging using nanodiamond quantum sensors enhanced by machine learning. *Sci Rep* **12**, 13942 (2022).

36. Wang, G. *et al.* Fast Wide-Field Quantum Sensor Based on Solid-State Spins Integrated with a SPAD Array. *Adv Quantum Tech* **6**, 2300046 (2023).

37. Troise, L. *et al.* High-speed magnetic-field imaging via N- V ensemble and laser raster scanning. *Phys. Rev. Applied* **25**, 064065 (2026).

38. Reed, N. R. *et al.* Machine learning for improved current-density reconstruction from two-dimensional vector magnetic images. *Phys. Rev. Applied* **23**, 034035 (2025).

39. Simpson, D. A. *et al.* Magneto-optical imaging of thin magnetic films using spins in diamond. *Sci Rep* **6**, 22797 (2016).

40. Simpson, D. A. *et al.* Magneto-optical imaging of thin magnetic films using spins in diamond. *Sci Rep* **6**, 22797 (2016).
41. Le Sage, D. *et al.* Optical magnetic imaging of living cells. *Nature* **496**, 486–489 (2013).
42. Tetienne, J.-P. *et al.* Quantum imaging of current flow in graphene. *Sci. Adv.* **3**, e1602429 (2017).
43. Chen, Y., Li, Z., Guo, H., Wu, D. & Tang, J. Simultaneous imaging of magnetic field and temperature using a wide-field quantum diamond microscope. *EPJ Quantum Technol.* **8**, 8 (2021).
44. Glenn, D. R. *et al.* Single-cell magnetic imaging using a quantum diamond microscope. *Nat Methods* **12**, 736–738 (2015).